\documentclass[10pt, conference, compsocconf]{IEEEtran}
\IEEEoverridecommandlockouts
\usepackage{graphicx}
\usepackage{listings}
\usepackage{color}
\definecolor{gray}{rgb}{0.4,0.4,0.4}
\definecolor{darkblue}{rgb}{0.0,0.0,0.6}
\definecolor{cyan}{rgb}{0.0,0.6,0.6}

\lstdefinelanguage{XML}
{
	morestring=[b]",
	morestring=[s]{>}{<},
	morecomment=[s]{<?}{?>},
	stringstyle=\color{black},
	identifierstyle=\color{darkblue},
	keywordstyle=\color{cyan},
	morekeywords={xmlns,version,type}% list your attributes here
}

\begin{document}
\title{In-situ data analytics for highly scalable cloud modelling on Cray machines}

\author{\IEEEauthorblockN{Nick Brown, Michele Weiland}

\IEEEauthorblockA{EPCC, University of Edinburgh, Edinburgh, UK}

\and

\IEEEauthorblockN{Adrian Hill, Ben Shipway}

\IEEEauthorblockA{UK Met Office, Exeter}

}

\maketitle

\begin{abstract}
MONC is a highly scalable modelling tool for the investigation of atmospheric flows, turbulence and cloud microphysics. Typical simulations produce very large amounts of raw data which must then be analysed for scientific investigation. For performance and scalability reasons this analysis and subsequent writing to disk should be performed in-situ on the data as it is generated however one does not wish to pause the computation whilst analysis is carried out.

In this paper we present the analytics approach of MONC, where cores of a node are shared between computation and data analytics. By asynchronously sending their data to an analytics core, the computational cores can run continuously without having to pause for data writing or analysis. We describe our IO server framework and analytics workflow, which is highly asynchronous, along with solutions to challenges that this approach raises and the performance implications of some common configuration choices. The result of this work is a highly scalable analytics approach and we illustrate on up to 32768 computational cores of a Cray XC30 that there is minimal performance impact on the runtime when enabling data analytics in MONC and also investigate the performance and suitability of our approach on the KNL.
\end{abstract}

\begin{IEEEkeywords}
Parallel processing; Multithreading; Software performance ; Supercomputers ; Numerical simulation ; Data analysis 
\end{IEEEkeywords}

\section{Introduction}
The Met Office NERC Cloud model (MONC) \cite{easc} is an open source high resolution modelling framework that employs large eddy simulation to study the physics of turbulent flows and further develop and test physical parametrisations and assumptions used in numerical weather and climate prediction. MONC replaces an existing model called the Large Eddy Model (LEM) \cite{lem} which was an instrumental tool, used by the weather and climate communities, since the 1980s for activities such as development and testing of the Met Office Unified Model (UM) boundary layer scheme \cite{lock1998}\cite{lock2000}, convection scheme \cite{petch2001}\cite{petch2006} and cloud microphysics \cite{abel2007}\cite{hill2014}.  

The simulations that these models run generate a significant amount of raw data, it is not this raw data itself that the scientists are most interested in but instead higher level information that results from analysis on this data. Previous generations of models, such as the LEM, which exhibited very limited parallel scalability were able to perform this data analysis either by writing raw data to a file and analysing offline, or by doing it in-line with the computation without much impact on performance. However as modern models, such as MONC, open up the possibility of routinely running very large simulations on many thousands of cores, for performance and scalability it is not possible to write this raw data to file and do analysis off-line or stop the computation whilst analysis is performed in-line. This situation is likely to become more severe as we move towards exa-scale and run these models on hundreds of thousands of cores.

In this paper we introduce the data analysis framework approach and implementation that we have developed for MONC where, instead of computation, some cores of a processor run our IO server and are used for data analysis. The computation cores ``fire and forget" their raw data to a corresponding IO server which will then perform the analysis and any required IO. In order to promote this ``fire and forget'' approach, where computational cores can be kept busy doing their work, the IO server is highly asynchronous and has to deal with different data arriving at different times which raises specific challenges. After discussing the context of MONC and related work by the community in more detail in section 2, section 3 then focuses on our IO server, the analytics workflow and specific challenges we face in order to support scalable and performant analysis. In section 4 we introduce the active messaging abstraction that has been adopted to aid with the uncertainty of data arrival and ordering, before discussing collective writing optimisations in section 5. Performance and scaling results of our IO server running a standard MONC test case on up to 32768 computational and 3277 data analytics cores of a Cray XC30 are presented in section 6, as well as performance results on a KNL Cray XC40, before drawing conclusions and discussing future work in section 7.

\section{Background}
MONC runs over many thousands of cores \cite{easc}, making it possible to model the atmosphere at scales and resolutions never before attainable. However these much larger simulations result in much larger data set sizes and the global size of the \emph{prognostic} fields (directly computed raw values, such as pressure or temperature) is often TBs at any point in time. These large prognostics are often not of primary concern to the scientists but instead \emph{diagnostic} fields, which are values resulting from data analysis on the raw prognostic data, are far more useful. One such diagnostic might be the minimum cloud height and the model proceeds in timesteps with samples taken from the prognostic fields every timestep, analysed, and then averaged over a specific time frame to produce the final diagnostic values. For instance every 5 minutes of model time we might output the average lowest cloud height over that five minute period, this is an example of time averaging each specific contribution. In MONC data analysis therefore includes direct analysis on the prognostic fields (which produces higher level diagnostic fields), and also time manipulation of both diagnostic and prognostic fields.  

One approach would be to write out the entirety of the raw prognostic fields to disk and then perform analytics offline. However for large simulations this could involve writing out hundreds of GBs every timestep, and much of the analytics requires contributions from every timestep. Not only would this post processing approach require very significant amounts of space on the community filesystem, as the computational runtime of a timestep is measured in milliseconds then the additional cost of IO that would be incurred is likely to be very significant. Previous generations of these models, such as the LEM, utilised an in-line approach to data analytics where cores would do both computation and analytics on the data as it was generated. However the LEM was not able to scale beyond 16 million grid points over 192 cores, in contrast MONC has been used to model systems of over 2 billion grid points on 32768 cores, with the plan being that it will run on over 100k cores in the future. As data and parallelism starts to reach such a scale it is important that the computation runs continuously. Due to the data size and cost of IO we endeavour to perform our analysis in-situ, as the data is generated, but don't want the computational cores to pause whilst doing this in-situ analysis themselves. Instead we wish to offload the data as it is generated to an IO server that will take care of data analysis and writing so that the computational cores can continue working on the next timestep.

MONC is written in Fortran 2003, uses MPI for parallelism and NetCDF for IO. The target users are scientists in the climate and weather communities, these users often want to add in specific data analytics, but they often do not have an in-depth HPC programming background. Therefore it is important to support powerful configuration, easy extendability, and use existing technologies (such as Fortran) which the users are already familiar with, not only so they can modify or add functionality themselves but also because these technologies are known to scale well and are supported by their existing software ecosystem. 

\subsection{Data analytics requirements}
Driven by the planned use of MONC, in addition to the community itself, we realised that there are a number of requirements that the data analytics approach adopted must implement. Many of these raise specific challenges that we discuss, along with our solutions, later in the paper in more detail. It is useful to be aware of these specific requirements at this point as it provides context to the related work discussed in section \ref{sec:related}.
\begin{itemize}
	\item \textbf{Dynamic time-stepping}: The MONC computational model proceeds in dynamic timesteps, where the exact size of a timestep varies depending on the computational stability of the system. It is therefore not possible to predict in advance what data will exactly represent because its origin point in model time is unknown until arrival. This adds considerable complexity to the handling of data, not least because we don't know the size of numerous output file dimensions until final write time. Details about how we solve the challenges that a dynamic timestep raises is discussed in section \ref{sec:architecture}.
	\item \textbf{Checkpoint-restart of the IO server itself}: Whilst it is common to checkpoint and restart computational models, for instance for long running jobs, with MONC we also need to be able to checkpoint and restart the state of the data analytics. Due to the asynchronicity that we promote (see section \ref{sec:architecture}) it is challenging to get an entire, reliable, snapshot of the data analytics state at a specific point in time. These challenges and our solution to them is discussed in more detail in section \ref{sec:checkpointing}.
	\item \textbf{Bit reproducibility}: The output from MONC runs need to be bit reproducible which requires that specific floating point operations are performed in a predictable and consistent order. This is discussed more in section \ref{sec:bitrepo}.
	\item \textbf{Scalability and performance}: Designed to be run on modern Cray machines, the MONC model is designed to simulate large domains over many thousands of processor cores. Whilst we accept that there is generally a runtime cost to enabling additional functionality, it is important that data analytics does not have a significant impact on the performance or scalability of the model. This is discussed more in section \ref{sec:performance}
	\item \textbf{Easy configuration and extendibility}: It is important that scientists can easily configure existing and write new diagnostics. Additionally some users will wish to modify the code itself to add or modify more complex functionality. The predominant language in use by the weather and climate community is Fortran, hence if we are to ensure the accessibility of this code to the community then that is the technology which must be used. The challenges of configuration and approach we adopted is discussed in section \ref{sec:userconfig} and extending the IO server in section \ref{sec:extending}.
\end{itemize}

\subsection{Existing approaches}
\label{sec:related}
The XML IO Server (XIOS) \cite{xios} is a library dedicated to IO that was initially developed with the requirements of climate and weather codes in mind. Cores can also be dedicated to IO and service the computational cores of the model. Coupled with numerous models such as the NEMO ocean model, this C++ library allows one to perform IO in an asynchronous manner. The central idea is that at each timestep computational cores will expose some or all of their data to the IO server through a minimalist interface which is then processed depending upon the user's XML configuration. From \cite{xios-tut} it is clear that utilising XML is a clean and accessible method of configuration and XIOS defines many different actions that can be performed on data such as filters, transformation and basic numerical operators. The authors of XIOS view this asynchronous approach as a software IO burst buffer and this can smooth out the overall cost of IO. We have previously experimented with integrating XIOS in with MONC, however at the time XIOS did not support dynamic time-stepping (where the difference in timestep size changes continuously for numerical stability) which is an important part of our model and adds significant uncertainty and complexity to some of the analysis as discussed in section \ref{sec:architecture}. Whilst the general design of XIOS and mode of configuration are very convenient, the general nature of this approach inevitably means that there are facets which are not required by MONC. A prime example of this is the calendar, which is an important concept for XIOS but not relevant in MONC. Whilst it is possible to checkpoint the current state of the computational cores, it is not clear from the user guide \cite{xios-ug} whether one can checkpoint the state of XIOS itself and reinitialise XIOS based upon this. This checkpoint restart, not just of the computational cores but also data analytics, is an important aspect of the MONC model.

Damaris \cite{dorier:hal-00715252} is middleware for I/O and data management and, similar to XIOS, targets asynchronous IO by dedicating specific cores of a processor for this purpose. A major advantage is that Dameris is integrated with other common HPC tools such as the VisIt visualisation suite which enables in-situ visualisation of data. Written in C++ it can be extended through plug-ins written in a variety of languages, including Fortran and Python, and has mature Fortran bindings. Similarly to XIOS it is configured by XML but is currently very much at the framework stage, where analytics functionality would need to be explicitly developed and plugged in for different domains and it is a shame that a library of existing plug-ins is not available. From the user documentation \cite{damarius-ug} it is unclear how easy it would be to combine multiple plug-ins dynamically, for instance MONC requires specific analytics that involves not only a number of activities but also field values need to be collated and manipulated over time. Time manipulation raises specific issues around bit reproducibility, as discussed in section \ref{sec:bitrepo}, and it is unclear whether Damaris could abstract this or whether the specific plug-ins would need to handle it explicitly. It is also unclear whether the framework could perform checkpoint-restarts of its plug-ins or whether these would require explicit support.

The Adaptable IO System (ADIOS) \cite{liu2014hello} aims to provides a simple and flexible way for data in code to be described that may need to be written, read, or processed external to the computation. Configuration is done via XML files and there are bindings for a variety of languages. However ADIOS is focused very much around the writing of data rather than performing analytics on the data as it is generated. There are some data transformations provided which transparently modify the data on the fly during the write phase, but these are focussed around aspects such as compression rather than analysis. An interesting aspect of this technology is the fact that for performing the physical write a number of different transport methods are implemented and the user can easily configure which method(s) to use. A number of experimental transport methods are also available which have been developed in order to utilise less mature IO technologies. As a write layer ADIOS looks like a very interesting technology, however data analytics is not present and as such would only fulfil a fraction of our requirements.

The Unified Model IO server \cite{um-io} is a threaded server which utilises some of the cores of a system in order to perform IO. It provides a number of configuration parameters and has exhibited good performance in some cases \cite{um-io}. However there are some limitations, for instance IO is serial and routed through an individual process which will significantly increase memory requirements and limit performance. It doesn't support out of order execution either, which they note in \cite{um-io2} is a limitation and means data must be processed and written in a strict order. Interesting in \cite{um-io} they note that running under \emph{MPI\_THREAD\_MULTIPLE} mode provides the best performance in contrast to other MPI threading modes.

It can be seen that the technologies described in this section make heavy use of XML for user configuration and the concept of sharing cores between computation and IO/analytics. Specific technologies, such as Damaris, view themselves as a framework which users can extend and plug their specific functionality into. However non of these technologies, in their current state, support the requirements of the MONC model. Specific aspects in MONC that are lacking in these technologies are support for dynamic time-stepping, the requirement to checkpoint and restart based on current IO state and seamless bit reproducibility. Whilst it might seem logical to take one of these existing technologies and extend it to support these required features, this would be a considerable undertaking, require significant re-engineering of the technology and involve design issues around backwards compatibility. There is also the user requirement that the code be written in Fortran to reduce the barrier of extending the IO server by the community in future.

\section{MONC in-situ data analytics}
\begin{figure}
	\begin{center}
		\includegraphics[scale=0.5]{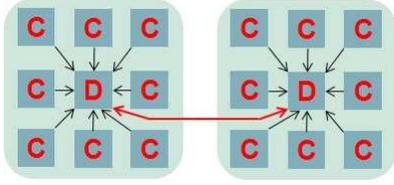}
	\end{center}
	\caption{Computation and data analytics cores}
	\label{fig:datacomp}
\end{figure}

In the previous section it was seen that no existing IO and analytics technology met all the requirements of MONC but many of the design decisions already proven could be reused. By designing an IO server and analytic approach specifically based upon the requirements of MONC users the aim was to provide an implementation that can be relied upon by scientists and compliment the mature computational model at scale. In our approach the cores within a node are shared for both computation and data analysis, as per figure \ref{fig:datacomp}, where typically one core per processor will perform the analysis (marked D, running the IO server) for the remaining cores which perform computation (marked C.) Much of the analytics involves combining local values to form some final result which is therefore more communication and IO bound rather than computation bound. In these situations a data analytics core (running our IO server) will compute contributions from its local MONC computational cores before communicating with all other data analytics cores to produce the final result. Determined by the user's configuration, at any timestep a group of raw prognostic fields may be communicated from a MONC computational core to its IO server. 

\subsection{User Configuration}
\label{sec:userconfig}
The IO server is self contained and initialised from a user's XML configuration file which controls both the behaviour of the server and also the high level data analytics to be performed. A typical XML IO configuration file contains three main parts, the first section defines groups of fields that will be sent from the computational cores to MONC and this is illustrated in listing \ref{lst:xmlconfigone}. Each group of raw data fields is given a name (in this case \emph{raw\_fields} ) and a timestep frequency at which the computational cores will send the data to its corresponding IO server. Each group is made up of a number of individual raw fields, in this case fields \emph{w} and \emph{u} are 3 dimensional arrays of doubles in sizes \emph{z} by \emph{y} by \emph{x}, each MONC computational core will contribute different parts of the field (hence \emph{collective=true}) and it is optional for these to be included in the group or not (hence \emph{optional=true}). Explicit field sizes are optional, for instance field \emph{vwp\_local} is a one dimensional array but the data size is not defined ahead of time so will be inferred at runtime.

\begin{lstlisting}[frame=lines,caption={XML configuration for receiving raw data},label={lst:xmlconfigone}]
<data-definition name="raw_fields" frequency="2">
  <field name="w" type="array" data_type="double" size="z,y,x" collective=true optional=true/>
  <field name="u" type="array" data_type="double" size="z,y,x" collective=true optional=true/>
  <field name="vwp_local" type="array" data_type="double" optional=true/>
</data-definition>
\end{lstlisting}

Listing \ref{lst:xmlconfigtwo} illustrates the second section of a typical configuration file and this part controls the data analytics of raw fields received from computational cores. In our example we have defined data analytics to be performed on the \emph{vwp\_local} raw field, generating the final analysed field, \emph{VWP\_mean}. In order to perform this work a number of rules will be executed, each transforming the raw data as configured by the user to intermediate or final states. Upon start up the IO server will read in these rules and perform dependency analysis, based upon intermediate values, to determine which rules must be executed before others and which can run concurrently. For example in listing \ref{lst:xmlconfigtwo} the first \emph{localreduce} operator rule (for summing up all column local values into one single total sum scalar) will execute first as this generates an internal variable, \emph{VWP\_mean\_loc\_reduced}, which is a dependency for the next rule. This second rule utilises the intermediate locally reduced values and issues a \emph{reduction} communication to sum up these scalar values across all IO servers. There are two parts to this type of communication, where it will first sum up all local contributions (made by MONC computational cores serviced by this IO server) and then communicate with other IO servers for the global value. A root process is provided which will be sent the resulting value, \emph{VWP\_mean\_g}. In the example this is specified as \emph{auto} and this means that the IO servers will determine the root at runtime and balance out roots between rules to aid with load balance. At this stage only the root has a value and other, non-root, IO servers will cease executing the analysis for the field. The root then activates the third rule, invoking the arbitrary arithmetic operations on the data, which in this case averages the summed up value to produce a single, final averaged \emph{VWP\_mean} variable. If a user specifies the \emph{units} decorator then this meta-data is provided in the NetCDF file in addition to the diagnostics value itself.

The \emph{arithmetic} operator rule in listing \ref{lst:xmlconfigtwo} also illustrates another design decision we have made in configuring the data analytics. It is intended that XML configuration is write once for a particular type of run and decoupled from the specifics of model execution and configuration. For instance different domain sizes, termination time other specific MONC computation configuration options should not require manual modification of the XML configuration. The \emph{\{x\_size\}} and \emph{\{y\_size\}} literals are an example of this, where any value enclosed in braces will be substituted by the IO server for specific MONC configuration options. For instance \emph{x\_size} and \emph{y\_size} are defined in a MONC model run configuration to determine the size of the domain in the horizontal dimensions, in this case these values are substituted in and used during the arithmetic operation. It means that for a specific run there is a single point of truth (the overall MONC computation configuration file), rather than the user having to remember to modify duplicated options in multiple places.

\begin{lstlisting}[frame=lines,caption={XML configuration for data analytics},label={lst:xmlconfigtwo}]
<data-handling>
  <diagnostic field="VWP_mean"  type="scalar" data_type="double" units="kg/m^2">
    <operator name="localreduce" operator="sum" result="VWP_mean_loc_reduced" field="vwp_local"/>
    <communication name="reduction" operator="sum" result="VWP_mean_g" field="VWP_mean_loc_reduced" root="auto"/>
    <operator name="arithmetic" result="VWP_mean" equation="VWP_mean_g/({x_size}*{y_size})"/>    
  </diagnostic>
</data-handling>
\end{lstlisting}

Listing \ref{lst:xmlconfigthree} illustrates the final part of a configuration file which involves writing both single and groups of fields to files. In this example a group of fields to be writen is initially defined, \emph{3d\_fields}, which contains the \emph{w} and \emph{u} fields. This grouping is for convenience as many different fields are often logically grouped and handled together, and one doesn't want to have to explicitly specify them many times throughout the configuration. An output file, \emph{profile\_ts.nc} is defined to be written every 100 model seconds, containing both the \emph{3d\_fields} group of (prognostic) fields and the \emph{VWP\_mean} analysed (diagnostic) field. All fields can be manipulated in time, either averaging over a specific model time period and producing the average value at a configured model time frequency or writing out an instantaneous snapshot at specific model time frequencies. This is defined by the \emph{time\_manipulation} option and in listing \ref{lst:xmlconfigthree} the \emph{VWP\_mean} diagnostic field is averaged over time with a value being written out every 10 model seconds, representing an average of the field values over that period. Conversely a snapshot of the \emph{w} and \emph{u} fields is written to the file every 5 model seconds, with every other field value being discarded.

\begin{lstlisting}[frame=lines,caption={XML configuration for writing to file},label={lst:xmlconfigthree}]
<group name="3d_fields">
  <member name="w"/>
  <member name="u"/>
</group>

<data-writing>
  <file name="profile_ts.nc" write_time_frequency="100" title="Profile diagnostics">
  <include field="VWP_mean" time_manipulation="averaged" output_frequency="10.0"/>
  <include group="3d_fields" time_manipulation="instantaneous" output_frequency="5.0"/>	 
  </file>
</data-writing>
\end{lstlisting}

Whilst we have intended these rules to be as simple and abstract as possible, they still require some high level knowledge and understanding of the data and architecture involved, which is not necessarily realistic for novice users. As such functionality to include other IO XML configuration files has been implemented. A large number of predefined analysis XML snippets have been implemented which, for instance, perform analysis for different types of field or write out values in different ways to the output file. Therefore many MONC users can simply prepare their configuration by importing these different snippets, with very little actual configuration having to be written explicitly by themselves. One of the challenges here was to ensure that snippets of predefined and user defined configurations do not conflict; specifically on the names of fields, variables, groups and what analysis rules should apply where. To this end it is possible to specify an optional namespace, where independent fields of identical names can co-exist in different namespaces and analytics, along with the writing of values to file, can explicitly specify a namespace to ensure encapsulation and avoid conflict. In the examples introduced in this section we have omitted a namespace which defaults to the global namespace. 

For each constituent rule contributing towards a specific diagnostic (as per listing \ref{lst:xmlconfigtwo}), the IO server identifies the corresponding activity implementation (the operator and/or communication) and forwards the XML arguments to this facet. It is the activity itself that is responsible for decoding and making sense of these arguments and-so we are not constrained to a set of centrally pre-defined hard coded arguments for all activities. Whilst a large proportion of analytics involves reductions there are a number of other forms required such as spectral methods and the tracking of clouds through the atmosphere. This, in combination with rule ordering enforced by dependency analysis, means that user's can develop complex analytics expressions. Generally speaking we see users falling into three general groups, the first are the experts who wish to deeply configure analytics and add new functionality. In combination with the XML configuration, the IO server has been designed such that new activities can be easily developed in Fortran and plugged in (see section \ref{sec:extending}.) The second category of users interact with the IO server at the configuration (XML) level only and rely on the existing activity implementations to achieve their analytics. Thirdly, novice users often wish to have minimal involvement with the implementation and in-depth configuration, we see the ability for them to import and specialise pre-defined configuration snippets as important.

\subsection{Analytics architecture}
\label{sec:architecture}
When the MONC model initially starts up each computational core registers itself with its corresponding IO server and at this point some handshaking occurs. IO servers send back information about the different fields that they expect from each computational core (based upon the first section of user's XML configuration) and the recipient computational core then responds with local information about these fields such as their size. This information is then used to build up MPI datatypes for each specific communication (group of fields per registered computational core) and buffer space is allocated so that computational cores can copy their data to a buffer, ``fire and forget'' this (via a non-blocking MPI send) to an IO server and then continue with the next timestep.

\begin{figure}
	\begin{center}
		\includegraphics[scale=0.35]{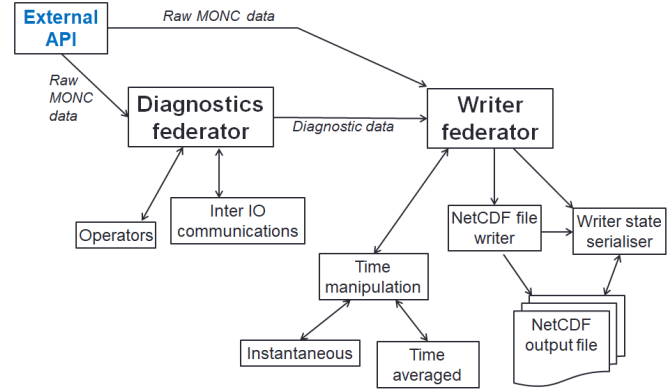}
	\end{center}
	\caption{Data analytics architecture}
	\label{fig:pipeline}
\end{figure}

Figure \ref{fig:pipeline} illustrates the IO server and analytics workflow, which is architected around a number of event handlers driven by two main federators. Each MONC computational core communicates with the IO server via an entry API. The IO server probes for external data messages, interprets these and, depending upon the message itself and corresponding user configuration, will then send this group of raw prognostic fields as an event to the diagnostics federator for data analytics and/or the writer federator for direct writing to file. A variety of functionality is performed by the federators and conceptually much of this is presented as event handlers sitting underneath a federator, awaiting specific events from the federator before activation. 

As discussed in section \ref{sec:userconfig} the second section of the user's XML configuration defines what data analytics the diagnostic federator should perform on what fields and the rules to execute in order to generate these diagnostics. A specific rule will execute when its dependencies, either prognostic fields or values generated from other rules, are available. Rules produce resulting values which are either the final diagnostic itself or an intermediate field that drives additional rules. These rules are made up of operators (such as arithmetic, field coarsening, field slicing and filtering) or communications (such as reductions or broadcasts.) They follow a standard template (Fortran interface) and are registered with the federator when the IO server initialises to match against specific rule names in the user's XML configuration. Operators for instance must implement procedures (via procedure pointers) for execution and determining whether the operator can run a specific rule or not. 

The writer federator in figure \ref{fig:pipeline}, defined by the third section of the users XML configuration as per listing \ref{lst:xmlconfigthree}, receives either an analysed diagnostic value from the diagnostics federator or raw prognostic fields directly from a computational core. Time manipulation is first performed which will either time average values over many timesteps or produce an instantaneous value at a specific preconfigured point in time. This time manipulation event handler will receive a field as an event, activate and perform any necessary computation upon it. It may or may not send an event back to the writer federator which would represent the time manipulated field and in which case the writer federator will store these values internally until they are physically written to disk. NetCDF-4 is used as the file format and IO technology, opening a file for parallel write by all IO servers at preconfigured specific points in model time, for instance every 100 model seconds. The output file contains both the field values themselves and meta-data such as the title of the file, model time, model timestep and field dimensions. Due to the time manipulation it is likely that there are many values for each field, each representing specific points in time, hence each actual field written contains an additional time dimension. Unique time dimensions are created for different user configuration setting and the file also explicitly specifies the time value at each point in the dimension. Whilst the NetCDF file writer is currently the only IO mechanism supported this has been defined in a general manner to allow for other mechanisms to be plugged in similarly to ADIOS.

In order to promote asynchronicity, which is important for performance and scaling, we avoid any synchronisation with the MONC computational cores and hence there is a high degree of uncertainty about exactly when data will arrive. But not only this; because MONC handles timestepping dynamically, where the exact size of a timestep varies depending on the computational stability of the system, we can not predict in advance what that data will represent because its origin point in model time is unknown. Therefore all groups of fields arriving at the IO server also include meta-data such as the current model time and timestep, this is checked and when it matches the write period a file write will be initiated which defines the NetCDF file and writes out stored values. Due to the asynchronicity, where one MONC core is ahead of another core, it is likely that not all required fields will be immediately available at the point of initiating writing; these are marked as outstanding and written when available. Another challenge is that the size of the timestep might become so large that it steps over multiple time manipulation periods, especially if these are configured to be quite short amounts of time. It is challenging to work with this uncertainty in a way that users can predict the behaviour of MONC, for instance if it is configured to write out every 100 model seconds we do not know exactly what the model time will be when the write is triggered and there is no way to predict this. For instance at one timestep the model time might be 98.2 and the next it could be 101.3, in this case encountering the latter the IO server determines that a write should occur and will write all values up to and including the configured write period, so in this case values representing 0 to 98.2 model seconds inclusive will be written and the values sent over at 101.3 model seconds will be stored for the next file.

In NetCDF opening and defining a file is collective, where each IO server must participate and define parallel fields collectively. However due to this possibility of jumping over time manipulation periods one can not be sure the number of entries in the time dimension until the explicit write point has been reached. Whilst the unlimited dimension would have helped with this issue, it is erroneous when writing in independent mode which we require. Therefore in our approach file definition does not begin until the write time has been reached and at this point each IO server has enough information to determine the number and size of dimensions required. Before writing, fields are stored in memory and when all fields have been written to the file and it is closed this memory is freed for re-use.

So far we have described each federator and event handler in a synchronous manner, such that it might be assumed that there is a single thread of execution. In reality each event handler is highly parallel and processes received events concurrent via threads. At its lowest level the IO server provides a shared thread pool and when an event is received by the handler it will request an idle thread from the pool and hand off the event to that thread for processing. As such very many events can be processed concurrently by the federator and, because the IO server is not computationally bound, we get good results when running with around a thousand threads in the pool. If there is no thread available then the handler will wait until notified by the pool that a thread is idle and can be utilised. Designing the architecture around the event based co-ordination pattern means that we can ensure that tricky aspects such as this will not result in deadlock. The \emph{Forthreads} \cite{awile2014pthreads} module is used wrap the pthreads C interface library which effectively gives us full pthread functionality in Fortran. We found that it was important to decorate variables shared between different threads with the \emph{volatile} attribute, which denotes that these might be modified by means not obvious from the code and hence not to rely on values held in cache. The \emph{volatile} attribute is one of the features of Fortran 2003 \cite{f2003} that we rely upon.

\subsection{Bit reproducibility}
\label{sec:bitrepo}
For simplicity the event handlers discussed in section \ref{sec:architecture} mainly process data (i.e. events) in the order in which they are received. This approach works well in many cases of the workflow, however can cause bit reproducibility issues where the event handler is accumulating field contributions such as with time averaging. Therefore specific aspects of the workflow, such as averaging field values over time, must be performed in a deterministic order due to the lack of associativity of floating point arithmetic. This relates to processing events in a deterministic order instead of the order in which they arrived. Because all data events in our approach contain meta-data, specifying aspects such as the timestep and model time, we can utilise this to determine the correct ordering. In cases where event ordering matters the handler will order based upon the timestep. From the user's configuration the handler is able to determine the timestep frequency of arrival for all data and hence the next expected timestep (but not the model time that this represents due to to dynamically timestepping.) Therefore events received out of order are stored in a queue until earlier events have been received and processed. In these event handlers, the processing of a specific event is therefore followed by checking the queue and then processing any appropriate outstanding events which can now be handled without compromising ordering constraints.

\subsection{Checkpointing}
\label{sec:checkpointing}
The MONC model can be long running, taking days or weeks of computational time, but often HPC machines have a runtime limit for jobs such as three hours on the Met Office XC40 (MONSooN) and twenty-four to forty-eight hours on ARCHER, an XC30. As such the model proceeds in episodes where MONC is run until a specific wall time, then writes out its state into a checkpoint file and will then restart from this checkpoint, which is known as a continuation run. For these long runs a script schedules two jobs, a job to run the model directly (either starting from initial conditions or a checkpoint file) and another job in a held state which is executed once the other job finishes. This second job will check whether a continuation run is appropriate and if so schedule another job in the held state dependent waiting for this job to complete and execute MONC with the latest checkpoint file.

Checkpointing the computational part of the model is fairly standard because a snapshot of the model for each core contains the raw prognostic fields and other data such as the model time, timestep and size of the timestep. However it is also necessary to snapshot and restart the state of each IO server core doing data analytics and this is more complex, not least due to asynchronicity. Messages from MONC computational cores or between IO servers can be in flight, waiting for additional local values before communication or in the process of being issued, therefore how to handle this reliably was a challenge. We rely on two facts, firstly that the majority of this non-determinism is contained within the diagnostics federator and its sub actions, and secondly based upon the timestep metadata associated with each field the writer federator can determine whether it has received all the expected fields up to a specific timestep or whether there are diagnostic or prognostic fields still outstanding.

When a checkpoint is triggered we therefore wait until the diagnostics federator has completed all of its work up until that specific timestep and made these fields available to the writer federator. At this point we only store the state of the writer federator, which is far more deterministic, and its sub activities such as the fields being manipulated in time and waiting to be written to file. The state of the writer federator can be split into five distinct areas (the writer federator and its sub activities, see figure \ref{fig:pipeline}) and each IO server will progress through each of the distinct states, serialising them into a stream of bytes for storage. Inside the checkpoint file each of these five states is represented by a specific variable which is written collectively by all IO servers. NetCDF parallel calls enable different processes to read or write at different offsets in the same variable concurrently, so all IO servers will write their own state at a unique point in the variable. Each variable is also associated with a directory of start points and byte lengths for each IO server rank so that on restart processes can easily determine exactly what they need to read in. Once a field is written to file the memory associated with serialising it is freed and the IO server will progress onto the next variable. 

Internally the state of a specific activity can be viewed as a tree, where nodes in this tree are sub-states. We have no way of knowing the size of these sub-states until they are examined and-so the only way of determining the overall memory required to serialise and hold a state is by walking its tree. Naively one might proceed by packaging the state's top level into memory, then reallocating memory for all sub-states and copying data between them, but we found this very slow due to the significant overhead that many memory allocations incurs. Instead the process of checkpointing proceeds in two phases for all activities, the first of walking all sub-states to determine the local memory size required. This is then reduced between the IO servers (with a sum) and used to collectively define the appropriate variables and dimensions in the NetCDF file. The second stage allocates the local memory chunk and physically packages up the state by serialising it into bytes ready for writing to the file. It is important that the state has not changed between the first and the second stage, or else the required memory size will be wrong. Therefore locks are issued during the first state of checkpointing and only released once a state has been serialised and packaged.

Another challenge with this approach was the global size of the serialised variables that are written into the NetCDF files. With many IO servers it is realistic that the size of the field will exceed 2GB and the Fortran interface to NetCDF only supports signed (hence 2GB rather than 4GB) 32 bit integers. This limits not just the dimension sizes, but also the specification of start locations and counts for parallel writes along with the integer data type. Instead we were forced to call directly to the NetCDF C interface via the ISO C bindings for storing and reading much of the IO server state to enable unsigned 64 bit integers, the standardised C interoperability mechanism is a feature of Fortran 2003 \cite{f2003} that we rely on. Whilst unsigned integers are more problematic in Fortran it would be useful if the developers of the NetCDF Fortran interface extended this to support long 8 bit integers at least.

\subsection{Extending the IO server}
\label{sec:extending}
The majority of users will interact with the IO server at the XML configuration level as discussed in section \ref{sec:userconfig}. However there are certain aspects where users, who are generally familiar with Fortran, might wish to create additional functionality and easily integrate this with the architecture. This comes back to the Damaris view, of an IO server as a framework which users can extend via their own code. An example of where this might be desirable in MONC are the operations that can be performed on data as part of analytics and as such as have defined them in a standard manner. Two Fortran interfaces have been defined which each operator must implement. The first determines which fields must be present to execute the operator for a specific configuration so that the diagnostics federator can determine dependencies and the ordering of rules for a specific diagnostics calculation. The second interface that must be implemented actually executes the operation and this subroutine takes as arguments the configuration of the IO server, the source id of the MONC computational core and the input field data itself. A central \emph{operator} handler (see figure \ref{fig:pipeline}) uses Fortran procedure pointers to map between the operator names and their underlying procedure calls. These procedure pointers are one of the features of Fortran 2003 \cite{f2003} that we rely on in the IO server implementation.

\section{Active messaging communications}
Allowing the MONC computational cores to fire and forget their data to a corresponding IO server is important to avoid interrupting the computation however this asynchronicity, where different IO servers can receive and process different data at different times, results in a specific challenge when it comes to data analytics. Collective communications between the IO servers is required for many diagnostic calculations, for instance to calculate global average, minimum or maximum values and whilst MPI version 3.0 provides non-blocking collectives, the issue order of these is still critically important. For instance if there are two fields, \emph{A} and \emph{B}, where one IO server issues a non-blocking MPI reduction for field \emph{A} and then the same for field \emph{B}, then every other IO server would also need to issue reductions in that same order. If another IO server was to issue reduction on field \emph{B} and then \emph{A} then the calculated values would be incorrect. One way around this would be for all the IO servers to synchronise before issuing reductions for specific fields to guarantee the issue order, but this would result in excessive communications, overhead and code complexity.

Instead we adopted the approach of active messaging where all communications are non-blocking and an IO server issues these, such as a reduction, and then continues doing other work or returns the thread to the pool. At some point, depending on whether this IO server is the root, a resulting message will arrive which will activate some handling functionality. Inside the \emph{inter IO communications} of figure \ref{fig:pipeline}, an IO server will call an appropriate active messaging function with common arguments such as the data itself and meta-data (such as data size and type), but will also specify a callback handling subroutine and string unique identifier. It is this unique identifier, instead of the issue order, that determines which communications match. For diagnostics the field name concatenated with the timestep number is generally used as the unique identifier. This is illustrated in listing \ref{lst:activemessagereduce}, where the non-blocking \emph{active\_reduce} subroutine is called for the \emph{VWP\_mean\_loc\_reduced} variable (as per the user's configuration of listing \ref{lst:xmlconfigtwo}.) This variable is a single floating point scalar on each IO server, which will be summed up and the result available on the root (in this case IO server rank zero.) The string unique identifier is also provided, along with the handling callback. At some point in the future a thread will call the \emph{diagnostics\_reduction\_completed} subroutine on IO server rank 0 with the results of this communication operation. The code in listing \ref{lst:activemessagereduce} is to illustrate this active messaging approach to the reader, in reality a generalised version of this has been implemented within the \emph{inter IO communications} functionality for active reductions of any size, message type and operation which is then called at runtime by the IO server depending upon user configuration.

\begin{lstlisting}[frame=lines,caption={Active messaging reduction example},label={lst:activemessagereduce}]
call active_reduce(VWP_mean_loc_reduced, 1, FLOAT, SUM, 0, "VWP_mean_loc_reduced_12", diagnostics_reduction_completed)
....

subroutine diagnostics_reduction_completed(data, data_size, data_type, unique_identifier)
....
end subroutine
\end{lstlisting}

Internal to the active messaging layer, that can be used throughout the IO server, unique identifiers and their associated callback subroutines are stored in a map. Built upon MPI P2P communications, when a message arrives it is decoded and passed into this layer for handling if it is deemed to be an active message. A look up is performed based on the unique identifier and, if found, a thread from the pool is activated to execute the callback with the message data and meta-data. In some cases it is possible that a message arrives on an IO server, but that IO server has not yet issued a corresponding communication call and hence no callback is found when it performs the look up. In these cases the data and meta-data is temporarily stored until such an API call has been issued locally. 

This approach is known as active messaging because it explicitly activates some handling functionality, running concurrently, based upon the arrival of messages. In this manner we need not worry about maintaining any message ordering, because this is done for us by the active messaging implementation based on the unique identifier. This also greatly simplifies the higher level data analytics code because one just needs to provide subroutine callbacks rather than deal with the tricky and lower level details. Abstracted from the analytics functionality lower levels of the IO server deal with issues around receiving data, checking MPI request handlers and executing the callbacks, rather than having to explicitly check request handles during analytics execution. In active messaging the registration of handlers can either be persistent, i.e. a registration will remain and activate for all message arrivals, or transient where registrations must match the arrival of messages and, once called, the callback is deregistered. In our approach the handlers are transient, the advantage of this is that error checking can be performed - for instance if a message is expected but never arrives.

\subsection{Barrier active messaging}
The active messaging approach has also proven very useful to avoid excessive synchronisation in other parts of the IO server code. NetCDF is not thread safe and as such we need to protect it with explicit mutexes, additionally when MPI is run in thread serialised mode we also need to protect  MPI calls using a different mutex. Parallel NetCDF uses MPI-IO and as such every NetCDF call in the IO server will not only lock out any other NetCDF calls until this has completed, but also any MPI communication calls if the IO server is running thread serialised. It is therefore desirable to minimise the time spent in NetCDFs call but a number of these calls, such as defining and closing files, are blocking collectives. In this case each IO server will wait until every other IO server has issued the same call which can be very expensive. Not only is the thread idle and not available in the pool for other work, but also whilst it is blocked no other NetCDF operations (on other files for instance) or MPI communication calls can be issued.

As the writing of files is driven by the arrival of data, which is asynchronous and non-deterministic, different servers will trigger IO operations at different times. From experimentation it was found that there was considerable drift here, where one IO server could be waiting for a substantial amount of time in this blocked state before other IO servers (which might have had more data analytics to perform or more computational cores to service) issued a corresponding call.

The barrier active messaging call can be used to address this, where threads execute the non-blocking barrier call with a unique identifier and callback function. At some point in the future, when all IO servers have executed this active barrier with the corresponding unique identifier, then a thread is activated on every IO server executing the callback subroutine which itself performs the NetCDF blocking call(s). This callback performs the actual NetCDF file writing operations such as definition or closing of the file. Because of the semantics of an active barrier being that a specific callback should be executed only once all IO servers reach a specific point, we can ensure that the blocking NetCDF calls are all called at roughly the same point in time and there is far less waiting in this blocked state.

\subsection{Termination}
The use of active messaging does make termination more complex. It is no longer enough that all MONC computational cores have completed, because outstanding analytics messages might also be in-flight. Therefore in addition to the computational cores finishing there are two other criteria that must be satisfied for termination. First that no event handlers are active, which we can easily find out by checking whether any threads in the pool are active. Secondly that there are no outstanding active messages to be handled. The transient nature of our active messaging layer makes this much easier, as we can perform a table lookup to ensure that there are no outstanding callbacks that have been registered in the active messaging layer but not yet activated. 

\section{Collective writing optimisations}
In addition to writing out diagnostic values, it is also sometimes desirable to write out the raw prognostic fields, for instance when checkpointing or for provenance. However these fields can be very large and not only is it important to do the writing collectively, so that each IO server will write out its specific contribution to the overall global field at the same time, but we also want to minimise the number of writes being performed. If each MONC computational core  was to do the writing of its prognostic fields itself then this would not be an issue because local prognostic data is contiguous, however with an IO server many computational cores might be sending their prognostic data for writing at different points and these represent different, potentially non-contiguous, chunks of the global domain.

\begin{figure}
	\begin{center}
	\includegraphics[scale=0.50]{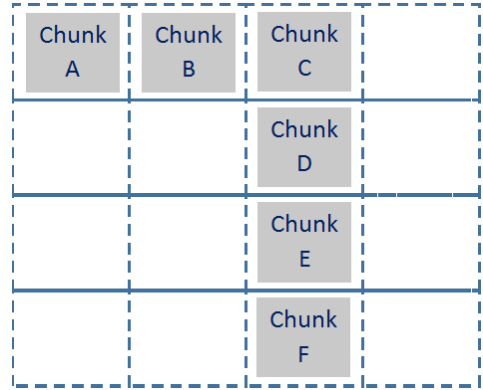}
	\end{center}
	\caption{Illustration of contributed data chunks in the global domain}
	\label{fig:datachunks}
\end{figure}

At start-up, once each computational core has registered with its corresponding IO server, a search is performed by each IO server over the sub-domain location of its MONC computational cores. The aim is to combine as many sub-domains as possible into contiguous chunks in order to minimise the number of writes. Because it is often the case that MONC computational cores of a processor are neighbours, working on geographically similar areas of the global domain, it is often possible to result in one or two large writes which is desirable. This is illustrated by figure \ref{fig:datachunks} where the algorithm will search in both dimensions and for each chunk will determine what larger contiguous chunks it could be a member of. Starting in the horizontal, the algorithm would identify that chunks A, B and C could form a contiguous block, but it will then progress to the vertical and determine that chunks C, D, E and F can form a contiguous block. Larger chunks are preferred over smaller ones, so in this case two regions (and hence writes) are selected, the first containing chunks A and B, and the second chunks C, D, E and F.

Memory space is allocated for these contiguous buffers and prognostic data from local computational cores are copied into the buffer which is then written to file once all contributions have been made. The writing of these fields is collective therefore all IO servers must participate for every write. A reduction with the \emph{max} operator is performed on initialisation to determine the maximum number of collective writes that any IO server will issue. Any IO servers with fewer writes than this will issue dummy (i.e. empty, of zero count) writes so that it is still involved in this collective operation.

\section{Performance and scalability}
\label{sec:performance}
\begin{figure}
	\begin{center}
		\includegraphics[scale=0.38]{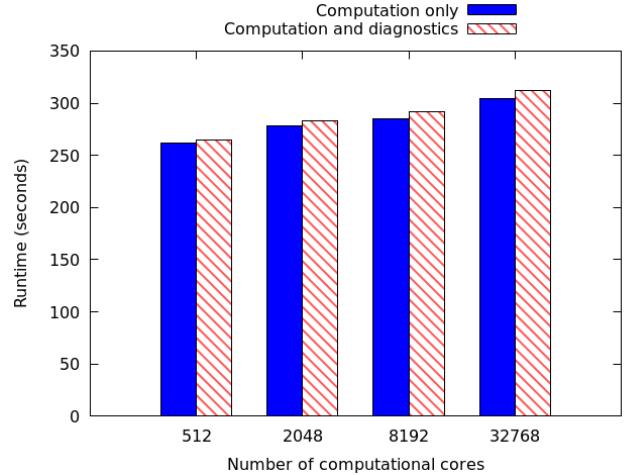}
	\end{center}
	\caption{Overall MONC runtime when weak scaling with and without data analytics, 65536 grid points per core}
	\label{fig:overallperformance}
\end{figure}

Performance and scalability tests have been carried out on ARCHER an XC30 (Ivy Bridge CPUs with the Lustre filesystem) which is the UK national supercomputer and one of the main targets for MONC and this in-situ analytics. A standard test case is used which is concerned with modelling stratus cloud in the atmosphere. 232 diagnostic values are calculated every timestep and time averaged with a result every 10 model seconds, the NetCDF file is written every 100 model seconds and the run terminates after 2000 model seconds. Figure \ref{fig:overallperformance} illustrates the performance of the model with and without data analytics at different core counts. In this experiment we are weak scaling, with a local problem size of 65536 grid points and on 32768 cores this equates to 2.1 billion global grid points. It can be seen that there is an impact of enabling data analytics of 8.14 seconds, or 2.6\%, at 32768 cores. An important point to note about this graph is that irrespective of whether analytics is enabled or not we have chosen to keep the number of computational cores the same. Therefore for the largest data analytics run, where one core per processor performs analytics, there is in fact a total of 36045 cores, 32768 for computation and 3277 for data analytics. Because we are increasing the overall number of cores there is a question of whether this is a fair experiment, but the number of computational cores remains the same and this represents the most common way in which MONC will be used. If we were to lock the number of overall cores and instead reduce the number of computational cores when enabling data analytics it would be more difficult to interpret the results as any increase in runtime could also be attributed to the fact that we have less computational power solving the problem.

\begin{figure}
	\begin{center}
		\includegraphics[scale=0.38]{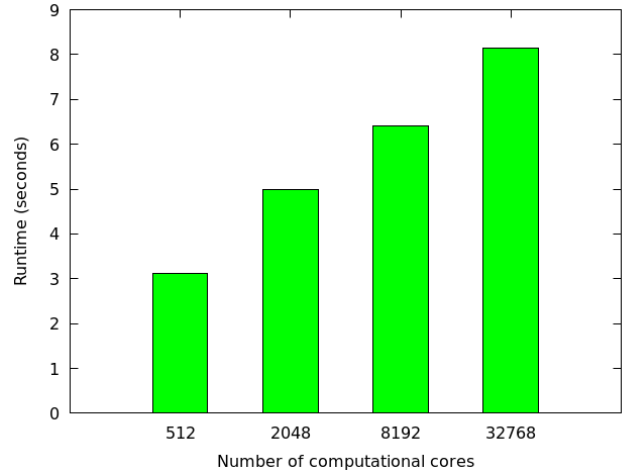}
	\end{center}
	\caption{Overhead of IO for different computational core counts weak scaling, 65536 grid points per core}
	\label{fig:iooverhead}
\end{figure}

In order to further examine the performance and behaviour of the IO server we have defined a performance metric which measures the elapsed time between a MONC computational process communicating a value at a specific model time that will induce a file write and that write then being completed. This is the overhead of data analytics and we want this time to be as small as possible, i.e. as soon as a write can be performed it is desirable for all the constituent data to be present and for this write to happen as quickly as possible. If the overhead of parallelism is too large then the IO servers might lag behind their computational cores which will result in excessive time at the end of the simulation whilst IO servers catch up. There are a variety of different factors that might influence the overhead, such as specific diagnostics not being completed yet, all threads in the pool busy so they can not action events in the system and the overhead of IO. Figure \ref{fig:iooverhead} illustrates the IO overhead for the weak scaling test case runs with 65536 grid points per core as we vary the number of computational cores. It can be seen that as one increases the number of computational cores the IO overhead also increases. This is because considerably more data is being processed globally and more IO servers are having to communicate during analysis and NetCDF file writing.

Table \ref{tbl:optimisation} illustrates the overhead of parallelism measure for data analytics over 32768 computational cores under a number of different configurations. It is accepted that for computational hybrid codes, running in MPI multiple threading mode is inefficient \cite{bull2010performance} but \cite{um-io} mentions that multiple mode is preferable for the UM analytics. The IO server supports both multiple and serialized mode, the latter where the IO server itself protects MPI explicitly. From this table it can be seen that running in serialized mode is a better choice for our IO server. This is somewhat of a surprise because the NetCDF layer uses MPI-IO which needs to be protected explicitly in serialized mode, but this is still preferable to the finer grained locking in multiple mode which incurs other overheads. Hyper-threading often does not provide a performance improvement in computationally intensive codes, but by keeping core placement the same (so it is unchanged for computational cores) and enabling hyper-threading, the thread pool of the IO server will automatically take advantage of these extra threads. It can be seen from table \ref{tbl:optimisation} that there is a performance benefit to using hyper-threading for the IO server both in seralized and multiple mode. It is believed that MPI thread multiple mode benefits from hyper-threading due to the finer grained locking involved. The experiments of figure \ref{fig:overallperformance} ran in serialized mode with hyper-threading enabled and it can be seen that the majority of the 8.14s runtime impact is due to this overhead of the last write at termination hence data analytics and IO is keeping up with the computational cores. Whilst some implementations of MPI do not support thread multiple mode, and some are known to be buggy, empirically it was found that Cray's implementation is stable and reliable with our IO server.

\begin{table}[h]
	\centering
	\begin{tabular}{ | c | c | }
		\hline
		Overhead (s) \quad&\quad  \\
		\hline			
		8.92 \quad&\quad MPI serialized mode\\
		12.02 \quad&\quad MPI multiple mode\\	
		\hline
		8.14 \quad&\quad Serialized mode + hyperthreading\\
		9.71 \quad&\quad Multiple mode + hyperthreading\\
		\hline
	\end{tabular}
	\caption{Configuration impact on overhead with 32768 computational cores}
	\label{tbl:optimisation}
\end{table}

\begin{figure}
	\begin{center}
		\includegraphics[scale=0.75]{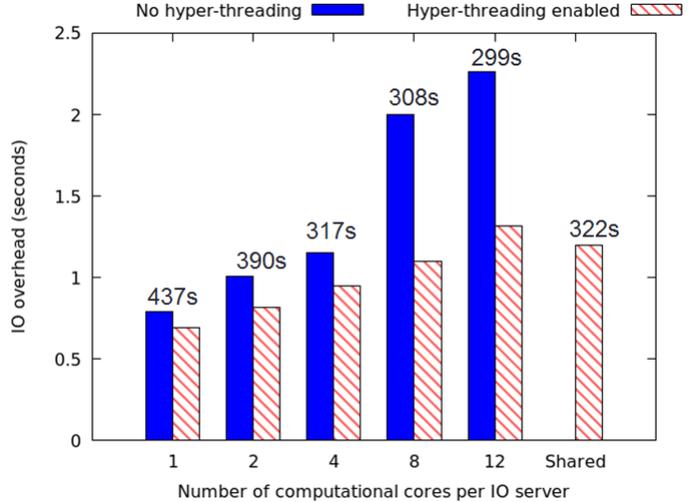}
	\end{center}
	\caption{Overhead of IO and overall test-case runtime on the KNL under different configurations with a global domain size of 3.3 million grid points}
	\label{fig:knl}
\end{figure}

Figure \ref{fig:knl} illustrates both the overhead of IO and overall model runtime for the same test-case on one Cray XC40 Knights Landing (KNL) machine (ARCHER KNL.) The node contains one 7210 Knights Landing (KNL) CPU with 64 cores, 16GB of MCDRAM run in quadrant cache mode and 96GB of main memory RAM attached to the Lustre filesystem. Each core is capable of up to four way hyper-threading and for this specific experiment we selected a global domain size of 3.3 million grid points. The results represent the overhead of IO and overall runtime as we varied the ratio of IO servers to computational cores which are mapped to physical cores, so for instance the ratio of one results in 32 physical cores for computation and 32 physical cores for IO. As one would expect from previous results in this section running with the four way hyper-threading that the KNL supports provides a significant reduction in overhead and we avoid the large spikes that a lack of hyper-threading for 8 and 12 computational cores per IO server exhibits due to having more physical concurrency to execute the threads processing data. It can be seen that, whilst the overhead of IO is minimal based upon a ratio of one, because we only have 32 cores dedicated to computation the overall model runtime is considerably longer than a ratio of 12 (59 computational cores and 5 IO servers.) 

The \emph{shared} column illustrates the situation where each KNL physical core runs MONC computation on one hyper-thread and the IO server on the remaining three hyper-threads, servicing its local computation. In this case it can be seen that performance is between that obtained when running with 8 computational cores to an IO server and 12. Beyond 12 computational cores to an IO server the overhead of IO grows exponentially, for instance with 32 computational cores per IO server the overhead of IO is around 500 seconds as there is just too much data for the threads to process concurrently, the IO server becomes swamped and gets significantly behind the computation. The conclusion we can draw from this result is that on the KNL one should endeavour to have as many computational MONC processes as possible by selecting a ratio of MONCs to IO servers that keeps up with computation but doesn't aim to optimise the overhead of IO. Beyond this point there will likely be a very sharp increase in both the overhead of IO and overall model runtime as the IO servers become swamped and one needs to find this optimal ratio empirically.

\section{Conclusions and future work}
In this paper we have described our approach and implementation of in-situ data analytics and file writing in MONC, where cores of a processor are shared between computation and data analysis. We have discussed some of the crucial aspects that this approach raises and lessons learnt such as the need for out-of-order collectives, which MPI does not fully fulfil, bit reproducibility, optimisation of large field writing and the ability to checkpoint analytics. We have experimented with configuration options and, by adopting our overhead runtime metric, we can see how some of the conventional wisdom that applies to the configuration of computation codes impacts data analytics. Runtimes for a common MONC testcase have been demonstrated on up to 32768 computational cores of a Cray XC30 and it has been shown that, for a typical configuration, our approach to data analytics and writing has limited impact on the runtime of the code. We have also examined the performance of our approach on a Cray XC40 KNL machine and illustrated the optimal configurations as well as raising the interesting idea of running one IO server per MONC computational core on the hyper-thread. Whilst Fortran is not a common technology for data analytics, it was used because this is what the computational code is written in and the scientists and their tools are familiar with it. We have shown that modern Fortran and associated library support is sufficient and our approach to data analytics is crucially important in order to support the science that the weather and climate communities will perform using MONC on current and next generation HPC machines. The IO server implementation is fairly specific to the MONC model and we believe that the major contribution of this work to the wider HPC community is based around the approaches adopted and lessons learnt that are detailed in this paper. As the MONC model is designed as a general purpose framework for atmospheric modelling there is some opportunity for direct re-use of the IO server within this context and at the time of writing scientists are planning on utilising MONC and the IO server for different forms of atmospheric science than the work was initially intended for.

In terms of further work, the active messaging layer which much of the diagnostics relies upon, is currently implemented on top of MPI P2P communications which is not necessarily optimal. Due to the fact that this is a lower level API then it would be transparent to the rest of the IO server to move this implementation closer to the lower level communications technology such as implementing it directly at the Cray DMAPP level or other similar technologies. We implemented the active messaging layer as part of the IO server because there was no suitable, mature, existing library that could be used. Previous research around active messaging has resulted in a number of technologies, for instance Charm++ \cite{charm}. Whilst Charm++ is mature and well supported, it requires the programmer to write their code in C++ and orient the parallelism around their concept of chars, which is not appropriate for the IO server. Other active messaging technologies such as AM++\cite{am} and AMMPI\cite{ammpi} are designed to be used in a more loosely coupled library call fashion, however these are all very much at the research stage and exhibit serious shortcomings including the lack of current development, limitations such as serial execution of the callbacks and requirement for specific synchronisations. In short we believe that the development of an active messaging layer, partly aided by lessons learnt during this work, would be of significant value not only to our IO server and in-situ data analytics in general but also more general codes as well.

The NetCDF file writer can be unplugged and other write mechanism technologies imported, it would be interesting to integrate with visualisation packages to enable in-situ visualisation of a simulation during execution. Whilst our IO server is currently fairly tied to the MONC model, it would be easy to extract it and make it more freely available for other models to utilise. 

Due to the community's familiarity with Fortran, up until this point the major focus has been on developing the entirety of the IO server in Fortran 2003 so that they can easily modify, extend and maintain the code. However moving forward it would be useful to support the integration of other languages specifically in terms of configuration. For instance in addition to the XML configuration that we have discussed in this paper it would be advantageous to enable users to write Python code snippets that also direct the analytics. Whilst the current configuration has been designed to be as expressible as possible, defining data analytics in a technology such as Python would increase accessibility for some people. Technologies such as f2py \cite{f2py}, which already ships as part of Numpy, support the integration of Python and Fortran codes so the complexities would be around the policy side, ensuring that integration is seamless, and performance, to understand whether there would be a runtime impact when using Python.

\section*{Acknowledgements}
This work was funded under the embedded CSE programme of the ARCHER UK National Supercomputing Service (http://www.archer.ac.uk)

\end{document}